\documentclass[11pt]{amsart}
\usepackage{geometry}                % See geometry.pdf to learn the layout options. There are lots.
\usepackage{graphicx}
\usepackage{amssymb}
\usepackage{epstopdf}
\title{Quantum Theory Needs (And Probably Has) Real Reduction}
                                     
\begin{document}
\maketitle

\centerline{R. E. Kastner}\smallskip
\centerline{\small {April 12, 2023; University of Maryland, College Park. rkastner@umd.edu}   }\bigskip

ABSTRACT. The traditional, standard approach to quantum theory is to assume that the theory ``really'' contains only unitary physical dynamics--i.e., that the only physically quantifiable evolution is that given by the time-dependent 
Schr\"{o}dinger equation. This leads to two distinct classes of interpretations for the standard theory in its orthodox form: (i) an Everettian-type approach assuming that all mutually exclusive outcomes occur in different ``branches'' of the universe; or (ii) single-outcome approaches that assume a ``projection postulate'' (PP) with no accompanying physical account within quantum theory. 

A contrasting, unorthodox approach is to suggest forms of quantum theory that involve physical non-unitarity; these are called ``objective collapse models.'' Among these are Penrose's theory of gravitation-induced collapse and the Transactional Interpretation. The primary focus of this paper is an example demonstrating that standard quantum theory (with or without the projection postulate) can in-principle yield empirically consequential inconsistencies. Thus, it appears that for quantum theory to be viable in a realist sense (as opposed to being an instrumentalist protocol in which inconsistencies are evaded by changing the protocol), it must possess genuine, physical non-unitarity yielding well-defined single outcomes. This leads to the conclusion that objective collapse models should be more seriously considered.

\section{Introduction}

It is a distinct honor and privilege to be able to contribute to this special issue commemorating the many accomplishments and contributions to physics of Sir Roger Penrose. The relevance of the present work is his advocacy of a form of quantum theory involving ``objective reduction'' or collapse, for which the present author can be viewed as a kindred spirit, despite our differing approaches to the specific form of such a theory.

First let us take stock of the basic ``lay of the land.'' The standard approach to quantum theory---what we might call the ``Received View''---is to assume that the theory ``really'' contains only unitary physical dynamics--i.e., that the only physically quantifiable dynamics is the deterministic evolution given by the time-dependent  Schr\"{o}dinger equation. This leads to two distinct forms for the standard theory: (i) an Everettian-type approach assuming that all mutually exclusive outcomes occur in different ``branches'' of the universe; or (ii) single-outcome approaches that assume a ``projection postulate'' (PP) with no accompanying physical account within quantum theory. The latter choice involves the so-called ``Heisenberg Cut,'' which partitions all physical systems as follows: a subset is taken as described by quantum theory, while its complement (which includes measuring instruments and/or observers) is taken as lying outside quantum theory. The PP is then invoked at the ``Cut'' to regain our empirical experience of a single outcome at the level of the latter set. Thus, the orthodox approach assumes that quantum theory itself, being solely unitary (and therefore deterministic), cannot yield a single outcome apart from the application of the PP, which involves a suspension of unitarity only through a suspension of the application of the theory.

There is yet a third class, involving some form of hidden variables (iii): this class comprises the ``retrocausal'' and ``superdeterministic'' interpretations which stipulate all past and future measurement outcomes (e.g., Andreoletti and Vervoort, 2022), as well as the deBroglie-Bohm theory (Bohm, 1952). Class (iii) may be considered unorthodox in that these theories involve adding hidden variables to the standard quantum theory. In superdeterministic/retrocausal theories, the stipulated measurement outcomes amount to hidden variables, since they are taken as enforcing single outcomes at specific times despite assumed ongoing unitarity as regards the quantum state. One might say that according to these theories, an evolving quantum state ``carries its measurement outcome(s) with it'' as an additional parameter that is generally unknown---hence, a hidden variable.\footnote{But see Kastner (2017) for an argument that there is no genuine dynamical ``causation'' in these kinds of ``retrocausal'' theories; they imply a static block world ontology.} 

In order to have a reasonably well-defined taxonomy, let us refer to all forms of the theory mentioned above as ``standard quantum theory'' (SQT). Classes (i) and (ii) are forms of orthodox standard quantum theory (OSQT), while (iii) is unorthodox standard quantum theory (USQT). They are all ``standard'' because they assume that ``real'' quantum theory has only unitary dynamics. The orthodox approaches eschew hidden variables, while the unorthodox approaches make use of hidden variables, either explicitly or implicitly, in various forms.

A contrasting approach, both unorthodox and {\it non-standard}, is to suggest forms of quantum theory that involve specific physical, quantifiable accounts of non-unitarity resulting in a single outcome upon measurement without the need for an arbitrary ``Cut''; these are called ``objective collapse'' or ``objective reduction'' models. Most of these approaches involve alteration of the Schr\"{o}dinger equation through additional nonlinear terms that serve to precipitate collapse to a single outcome. The models of Roger Penrose (1996, 2014) and Lajos Di\'{o}si (1987, 1989, 2013), as well as those explored by Ghirardi, Rimini, Weber (1986) and their followers, are in this class. Another objective collapse model, not involving alteration of the Schr\"{o}dinger equation, is the Transactional Interpretation (TI) originated by John Cramer (1986). TI is a different formulation of the standard theory using the direct-action theory of fields, so in that sense it is technically a different theory, although it is empirically equivalent to the standard theory at the level of the Born probabilities. TI contains physical non-unitarity by virtue of the behavior of the quantum electromagnetic field itself, rather than by changing the Schr\"{o}dinger equation. The present author has been developing the transactional picture into the relativistic domain in recent years, and it is now called the Relativistic Transactional Interpretation (RTI) in the literature (cf. Kastner, 2022, 2021a).\footnote{The relativistic version, RTI, differs empirically from the standard theory only in quantitatively predicting the non-unitary process of measurement (as formulated by von Neumann). It thus resolves an empirical anomaly---evidence consisting of single outcomes---for the standard theory.}

This work examines the implications of inconsistencies brought to light in recent years (and not-so-recent years, e.g. Deutsch, 1985) that afflict standard quantum theory (both orthodox and unorthodox as defined above). I will argue that these inconsistencies are fatally problematic, such that in order to make progress, it should be recognized that quantum theory needs (and in its fully correct form already has) real non-unitarity corresponding to reduction of the quantum state. While the present author has been studying a different non-unitary theory from that of Sir Roger, our common ground is recognizing the need for quantum state reduction. In brief, Sir Roger believes that Nature already provides this non-unitarity through gravitation, while I believe that Nature already provides it in the form of the direct-action theory of fields. 

\section{Background: Inconsistencies Facing Standard Quantum Theory}

Let us now turn to the issue of inconsistencies facing the standard theory, in particular those arising from Wigner's Friend-type scenarios. An examplar is the  Frauchiger-Renner paradox (Frauchiger and Renner, 2018), in which an application of the standard theory yields disagreement among different observers concerning the probability of a particular outcome.  Based on these and related paradoxes, some authors have argued that facts (i.e., measurement outcomes) are only relative and can be mutually inconsistent (e.g., Brukner, 2015; Proietti et al, 2019;  Bong et al, 2020; although see Kastner (2021b) for a critique of these claims). These presentations have assumed, either implicitly or explicitly, that such inconsistencies would never rise to level of empirically detectable inconsistency; i.e., that conflicting outcomes and/or or predictions could not be compared by the relevant parties. For example, Brukner (2015) argued that the inconsistent outcomes should not be considered to obtain together ``[a]s long as there is no communication on the relevant information (the actual measurement outcome) between the two laboratories'' (p. 22).   However, it has been pointed out by the present author (Kastner 2020b) as well as in Baumann and Brukner (2019) that such inconsistent outcomes can in fact be communicated, and are therefore more consequential than has been generally understood. I discuss herein a specific counterexample highlighting the fact that standard quantum theory can result in detectable--not just private and incommensurable--inconsistencies, which appears to constitute a {\it reductio ad absurdum} of the basic theory. I then consider the implications for various extant interpretations of quantum theory and argue that they do not succeed in evading the {\it reductio}. 

First, let us review what is meant here by ``Standard Quantum Theory'' (SQT). As noted in the Introduction, this term refers to the standard assumption that quantum theory ``really'' has only unitary evolution, given by the Schr\"{o}dinger equation, and that any non-unitary (if it is invoked at all) consists only of a ``projection postulate'' (PP) with no accompanying quantifiable dynamics. Thus, it must be emphasized that my critique applies not just to no-collapse formulation such as Everett's (class (i) above) but also to hidden variables approaches (class (iii)) and formulations that appeal to the PP (class (ii)). The PP is not real non-unitarity, as it contains no real physics. While the probabilities of putative outcomes arising from the PP can of course be described by the Born Rule, in the standard approach (SQT) there is no physical account of the transition from the pure state to the proper mixed state\footnote{A proper mixed state is not one obtained by `tracing over' degrees of freedom as in the usual decoherence approach. The latter yields an improper mixed state that cannot be legitimately interpreted as reflecting the existence of an eigenvalue outcome describable by the corresponding eigenstate.} needed to apply the Born Rule. This transition was first identified by Von Neumann and termed ``Process 1,'' and he noted that it could not be found within SQT itself. This situation -- the lack of a quantitative physical account of an interaction constituting ``Process 1,''  is of course the measurement problem of SQT. Hidden variables theories (class (iii), USQT) are claimed to solve the measurement problem through the stipulation of measurement results by way of those hidden variables. However, in view of their presumption of undisturbed unitarity evolution of the quantum state, these theories remain subject to the inconsistencies, which arise from the resulting general lack of well-defined Boolean spaces corresponding to sets of measurements.

 In the absence of a physical explanation for the PP, many adherents of SQT end up applying what I will call the {\it Unitary Outcome Assumption} (UOA). This is a methodological operating assumption of many applications of the standard theory that is generally uncritically presumed valid (although arguably it is not, as reviewed in Kastner (2020b, 2021b)).  The UOA  essentially defines an ``observer'' (or at least a system relative to whom an outcome may occur) as a degree of freedom $F$ whose states are correlated via unitary-only processes with those of other degree(s) of freedom of interest $A$, and asserts that a measurement outcome applicable to $A$ is obtained by the ``observer'' system $F$ based only on that correlation.  For example, the UOA is presupposed in Proietti {\it et al}, Baumann and Wolf (2018), and in Rovelli's Relational Quantum Mechanics (RQM) (Rovelli, 1996). I should hasten to add that the UOA is completely independent of whether the projection postulate PP is applied, i.e., whether the PP is assumed to apply as an objective matter (which involves a ``Cut''), or only privately as a description applied by a particular local observer, or not at all. The UOA simply takes for granted that a degree of freedom $F$, by virtue only of its correlation with another system $A$, obtained an outcome corresponding to an eigenvalue of some observable for $A$. 
 
 It should be emphasized that the UOA is not a new notion introduced by the present author and it is not something I would defend. On the contrary, my present purpose is to disclose and identify the often uncritically invoked UOA for critical evaluation.  Even if not explicitly stated, the UOA has been implicitly assumed in most of the more recent applications of SQT.\footnote{If UOA is rejected while retaining SQT, such rejection is usually based on decoherence arguments, as in Zukowski, M. and Markiewicz, M. (2021); the insufficiency of such arguments is briefly revisited in note 10.}  Before even getting to the inconsistency challenge, it should be noted that the UOA is already suspect in that it overlooks the fact that a mere correlation is not necessarily sufficient to single out a particular observable as the one being measured. For example, a unitary correlation resulting in a singlet-type state or $m=0$ triplet, \newline
 $|\psi\rangle_{\pm} = \frac{1}{\sqrt 2} [|\uparrow\downarrow \pm  \downarrow\uparrow]$ (even if created via a Hamiltonian employing a particular observable) fails to define a measurement basis, since these states can be written in terms of an infinite number of equally valid observables $\sigma_\theta$ for all $\theta$. Thus, even if it was one's intent to measure, say, $\sigma_z$, the resulting entangled state does not in fact physically single out $\sigma_z$. This emphasizes how inadequate a mere unitary correlation is for obtaining well-defined outcomes corresponding to eigenvalues of a particular, unique observable. For convenience, let us refer to this objection to the UOA as {\it UOA degeneracy}.  A further weakness of UOA is that it implies that measurement is occurring in situations that one would not want to count as measurements; for example, according to UOA, electrons in a singlet state should be ``measuring'' one another (which is also problematic in view of the degeneracy issue).

\section{Counterexample}

  We now show that the common assumption that  inconsistencies involving specific outcomes must always remain hidden, based on the idea that they are only observer-relative, is untenable via a specific counterexample. The example (similar to a scenario of Baumann and Brukner, 2019) demonstrates that attributing outcomes to entangled subsystems, even if designated as only ``relative'' to the subsystem and involving only ``subjective collapse,'' (as discussed by Baumann and Wolf, 2018), leads to empirical inconsistency.\footnote{Deutsch (1985) considers a scenario with some similarities, but presents it as an empirical comparison between the Everettian picture and the Copenhagen interpretation. The present work considers implications of this type of empirical disagreement that do not seem to have yet been fully appreciated.} 
   Assume, as in Proietti {\it et al}, that microscopic quantum systems such as atoms, molecules, and photons can act as ``observers'' whose correlations with other similar degrees of freedom constitute ``measurements'' (this is the UOA). Let $W$ and $F$ comprise  degrees of freedom subject to 2D state spaces, such as spin-$\frac{1}{2}$ atoms. Let us suppose that $F$ is ``measuring''  another  spin-$\frac{1}{2}$ system $A$, prepared in equal superposition of outcomes `up' and `down' along a direction $Z$, via a Stern-Gerlach device. $F$ comprises 2 degrees of freedom: $B$, acting as a pointer/memory, and $C$, for communication.  $C$'s possible states are `ground' and `excited,' where the latter would be triggered by a photon signal from $W$. While this is a thought experiment and does not depend on any particular experimental setup, one could think of $F$ as a micro-detector whose $B$ states are correlated with $A$ via an appropriate interaction Hamiltonian. $C$ remains in its initial unexcited state $|0\rangle$ at this stage. Thus, based on the idea that projection to an outcome eigenstate is relative to an observer, according to $F$ after the ``measurement''  of $A$ by $B$, $A$ the relevant degrees of freedom are either in the state 
      
      $$ |\Psi\uparrow\rangle_{FA} =   |\uparrow\rangle_A  \odot |\uparrow\rangle_B \odot |0\rangle_C\eqno(1)$$
      or
       $$ |\Psi\downarrow\rangle_{FA} =  |\downarrow\rangle_A  \odot |\downarrow\rangle_B \odot |0\rangle_C\eqno(2)$$
       \smallskip
       
\noindent with equal probability--i.e., they are in a mixed state.  On the other hand, according to $W$,  $A$ and $F$ end up in a pure Bell state with $F$'s communication degree of freedom $C$ along for the ride:

        $$ |\Psi \rangle_{FA} = |\Phi^+ \rangle |0 \rangle_C = \frac{1}{\sqrt 2} \bigl ( |\uparrow\rangle_A  |\uparrow\rangle_B
        + |\downarrow\rangle_A  |\downarrow\rangle_B \bigl ) \odot |0\rangle_C\eqno(3)$$
    
Now, let $W$ subject the $B+A$ system  to a measurement of the Bell observable for which the state $|\Phi^+\rangle$  is an eigenstate. The Bell observable measurement is somewhat involved, but there is nothing to preclude it in principle. $W$ could comprise as many bilevel systems as one wishes in order to allow for a pointer, signaling system, and any other degrees of freedom needed to implement the required correlations via interaction Hamiltonians, under the assumption (UOA) that all that is required for a ``measurement outcome'' (or outcomes in an Everettian approach) is a one-to-one correlation between eigenstates of an observable of the system being measured and the states of the putative ``observer'' (as assumed in Proietti {\it et al}).\footnote{The Bell observable is not uniquely required. Alternatively, one could arrange for $F$'s pointer $B$ to be anti-correlated with $A$'s putative outcomes, and have $W$ conduct a measurement of the total spin of the system. This would reveal the inconsistency since $W$ predicts that the total spin $J$ must always be found to be 1, while according to F, W has a 50\% chance of finding the result $J=0$, for the singlet state.} $W$'s measurement is accompanied by a signal to $F$ as follows. An outcome finding the state  $|\Phi^+\rangle$  results in a photon being emitted to $F$ to excite his communication degree of freedom $C$. This could be implemented by having $W$'s outcome registration involve excitation of a degree of freedom with a very short decay time such that it immediately de-excites in a collimated direction to $F$. According to $W$,  $F$ should receive that photon for every run of the experiment.  This makes the inconsistency manifest, since according to F, his probability of receiving the photon is only 1/2  given his description of the situation by the mixed state (1,2). We can make the inconsistency even more glaring by giving $F$ two communication degrees of freedom $C^+$ and $C^-$ corresponding to the outcomes of W's measurement corresponding to the states $|\Phi^+\rangle$ and $|\Phi^-\rangle$ respectively. $F$ predicts that $C^+$ and $C^-$ should receive equal numbers of photon signals, while W predicts that $C^+$ should receive all the photons and $C^-$ none of them. 
  
   The counterexample demonstrates that SQT, together with the UOA as it has been commonly assumed in the literature, cannot preclude empirically consequential--not just contextual and hidden--inconsistencies. These inconsistencies arguably amount to violations of the uncertainty principle, which precludes the possession of properties corresponding to eigenvalues of incompatible observables by a single system. That is, $A$'s prepared state is `up along $X$,' and $F$'s putative measurement is of $\sigma_z$ (if we charitably disregard UOA degeneracy). According to the uncertainty principle, obtaining an outcome for $\sigma_z$ makes $A$'s  $\sigma_x$ value maximally uncertain. But if $W$ treats $F$ as a degree of freedom correlated with $A$'s original prepared state (in conformance with the linearity of SQT), then there is nothing to prevent $W$ from confirming, via a Bell observable measurement of $A$ and $B$, $A$'s original prepared state and thus its $\sigma_x$  value with certainty (since according to SQT, $W$ must obtain  $|\Phi^+\rangle$ every time).  Clearly, the uncertainty principle does not allow this. But SQT cannot preclude it, and nothing stops $W$ and $F$ from communicating this discrepancy via auxiliary correlated degrees of freedom.\footnote{Matzkin and Sokolowski (2020) also argue that the usual Wigner's Friend inconsistency amounts to a violation of the uncertainty principle. The present example makes the violation more explicit and striking, since it presents both the conflicting pieces of information to the {\it same} observer ($F$).  One might be tempted to think that we could avoid the uncertainty principle violation by appealing to the degeneracy of the state  $|\Phi^+\rangle$ with respect to the individual spin direction basis. But going this route would also result in a contradiction, as follows: $A$'s value for $\sigma_z$ would have to be both maximally certain (via $F$'s purported outcome) and maximally indeterminate (via $W$'s Bell measurement,  invoking UOA degeneracy). So under either approach, $F$ is presented with a contradiction upon receiving the information from $W$: he must conclude either that his system violates the uncertainty principle, or that his system both has a sharp value and does not have a sharp value for $\sigma_z$.}
   
Above, I dissent from the escape from the uncertainty principle violation found, for example, in Brukner (2015). Brukner claims that no such violation occurs because the conflicting outcomes should not be combined, or allegedly ``do not co-exist.'' This claim is based on an  appeal to hidden variables, as follows: ``The trouble with the assumption that values for [two incompatible observables] coexist is that it introduces ``hidden variables,'' for which a Bell's theorem can be formulated with its known consequences...''. The problem with this evasion is that it is not an outside, additional ``assumption'' that the two values coexist; the theory (SQT) itself is what forces that, in particular by applying conflicting probabilities (based on the mutually inconsistent putative outcomes of $F$ and $W$) to the {\it same} degree of freedom ($C$). So, I question the prohibition on saying that the outcomes co-exist. Of course we don't {\it want} them to co-exist, because this results in obvious mayhem. But they {\it de facto} do co-exist according to SQT, because {\it the theory itself yields both outcomes and does not preclude their being communicated and compared}. Put differently, the prohibition on saying that both outcomes co-exist is much like making excuses for a bad meal at a restaurant, say a steak with chocolate sauce poured over it. (In case it is not clear, the restaurant represents SQT and the bad meal represents the mutually inconsistent outcomes.) Rather than return the dish to the kitchen, the diner says ``we should not be combining chocolate sauce with steak.'' No, of course we shouldn't, but that is what the kitchen served. Let's place the blame where it lies: on that which produced the faulty product, not the person to whom it was served.\footnote{Elouard {\it et al} (2021) similarly argue that the inconsistent results should not or cannot be combined because that would violate the uncertainty principle, placing the onus on the `diner' rather than on the `kitchen' that served the bad product. Their scenario involves specific assumptions that make the considered measurements experimentally incompatible, but those assumptions are not obligatory so this does not remove the in-principle conflict.}

Thus, it is {\it standard quantum theory itself} that yields the set of conflicting outcomes amounting to unacceptable hidden variables, not a meta-level researcher making an additional assumption or inference of ``combining'' them. No such additional inference is necessary. It is SQT itself that hands us the offending set of hidden variables, which arise precisely because of SQT's inability to define ``measurement'' in a physically sound and consistent manner. This point, that a set of incompatible outcomes resulting from taking unitary correlations as ``measurements'' (i.e., from the UOA) itself amounts to an assertion of hidden variables, is elaborated in Kastner (2021b).  
   
 The thought experiment presents an open conflict concerning the probability of a detectable state change in a specific degree of freedom accessible to both parties, as opposed to inconsistencies among ostensibly private and non-comparable results. Thus, the recourse to ``relativity of outcomes'' is insufficient to prevent an overt, empirical clash. One might attempt to evade the inconsistency by questioning whether microscopic objects would be able to ``apply quantum theory'' in this way, in order to block the $F$ state assignment. A reply is that if one is going to allow microscopic objects to ``obtain outcomes'' (the UOA), then such objects should also be able to attribute the corresponding quantum states to the systems they are ostensibly measuring. Arguably, raising this sort of objection consists of ``moving the goalpost,'' and the onus is on the objector to specify what sort of system qualifies as an ``agent for applying quantum theory.''  In any case, one can conceive the same thought experiment in a macroscopic form (essentially what is presented in Baumann and Brukner, 2019). One might appeal to decoherence effects (but see footnote 10 for why this tactic is overrated) to argue that such an experiment would be impractical, but there is no escape from the in-principle inconsistency facing SQT. 
 
 The ambiguity about what counts as an ``agent'' is yet another manifestation of  SQT's lack of a specific physical interaction qualifying as ``measurement.'' It is precisely this lack of a physical criterion for {\it measurement  yielding outcomes} that results in the prevailing (often uncritical) antirealist turn towards considerations about who and what qualifies as an ``agent applying quantum theory'' -- which constitutes an instrumentalist stance towards the theory. If only certain sorts of systems are allowed to ``apply quantum theory'' and represent outcomes with eigenstates that pertain only relative to themselves, then such state assignments can only be about subjective knowledge states and cannot describe anything in the world. Lest this be taken as suggesting that we should be led to such antirealism, the counterexample still shows that the putative instrument (SQT) is unreliable. Thus instrumentalist approaches such as Qbism do not remain unscathed, since the instrument (SQT) yields empirical-level inconsistencies. If instead quantum theory describes the world (i.e., if it is taken in a realist sense), then objects are simply described by it, and actually possess outcomes based on specific interactions; no agent designation is required in order for it to remain applicable and correct. But of course, what is lacking in SQT is a specific measurement interaction. We return to this point in the Conclusion.

\section{Relational QM Refuted}

The Relational Interpretation of Quantum Theory (RQM) was pioneered by Rovelli (1996). The basic postulate of RQM is stated by its adherents as follows:

``The physical assumption at the basis of RQM is the following postulate:  The probability distribution for (future) values of variables relative to $S$ depend on (past) values of variables relative to $S$ but not on (past) values of variables relative to another system $S^\prime$.''\footnote{Laudisa and Rovelli (2021). It should be noted, however, that this is not a ``physical assumption.'' It is a calculational, instrumentalist postulate that says nothing about any underlying physics. In contrast, Boltzmann's assumption that the large-scale laws of thermodynamics were based on the microscopic behaviors of atoms was a physical postulate.}\smallskip

Laudisa and Rovelli (2021) address a Wigner's Friend-type inconsistency pertaining to the 
 Schr\"{o}dinger Cat Paradox, as follows:\footnote{Laudisa and Rovelli downplay the usual measurement problem pertaining the the cat being in a superposition by appealing to decoherence. They assert: ``Hence, the answer to the question `Why don't we ever see cats that are half alive and half dead?' is: because quantum theory predicts that we never see this sort of things. It predicts that we see cats either alive or dead.''  But in fact, this is simply not the case. As first argued by Feyerabend (1962), discussed in Hughes (1989), and reiterated in Kastner (2020), orthodox quantum theory (SQT) predicts only systems and pointers in improper mixtures, which specifically precludes that `we will see cats either alive or dead'. Appealing to decoherence in SQT leads only to approximate vanishing of off-diagonals corresponding to interference; but even if perfect diagonalization obtained, that would not constitute a proper mixture required for the existence of outcomes. It also relies on an essentially classical `computational basis' at the outset, thus being a circular account of diagonalization with respect to the desired local observables. Thus, SQT is unable to rule out a pointer basis for the cat of `alive $+$ dead' and `alive $-$ dead' except by fiat, by defining preferred local observables and partitionings of the degrees of freedom making up `systems' vs `environment' (Kastner 2014). So in no sense does standard SQT ``predict that we see cats either alive or dead.''  A claim that the basic measurement problem for SQT, as illustrated by Schr\"{o}dinger's Cat Paradox, is solved based only on decoherence is simply untenable.}

\vspace{3mm}
``A problem, however, appears in quantum mechanics if we ask what the cat itself would perceive. Say the brain of the cat measures whether its heart is beating or not. The theory predicts that the brain will find either that it does or that it does not. In textbook quantum mechanics, this implies a collapse of $\psi$ to either $\psi^\prime$ or $\psi^{\prime\prime}$ [the two superposed states]. In turns, this implies that no further effects of interference between these two states will happen. And this contradicts the conclusion that interference effects, although small due to decoherence, are nevertheless real. This problem is resolved by RQM by the postulate above: the way the cat, as a quantum system, affects an external system, is not affected by the specific way the heart of the cat has affected its brain. That is, the state of the cat with respect to the external world does not collapse when a part of the cat interacts with another.'' (Laudisa and Rovelli 2021).

\vspace{3mm}

 Thus, RQM asserts that the cat, playing the part of $F$, should be able to predict correctly the future values of its own local system(s) by reference only to its local past outcomes, while its presumed unitary entanglement with the result of the world continues. But in the counterexample (presuming that $W$ is correct), the inner system $F$ {\it cannot} correctly predict the probability of a future event relative to itself---i.e., the probability of its degree of freedom $C$ becoming excited---based on its own past outcomes. This probability apparently depends instead on another system outside ($W$), contrary to the central postulate. Thus, it appears that the main postulate of RQM is refuted by the counterexample.

Laudisa and Rovelli comment that ``The central claim at the basis of RQM is therefore that: `different observers can give different accounts of the same set of events' (Rovelli 1996: 1643), and ``The only reality in RQM is given by events, which are the result of interactions between distinct quantum systems, but even these events can be described in a different way by different physical systems.'' (Laudisa and Rovelli, 2021). This approach is intended to be analogous to the differing descriptions of the invariant set of events in relativity based on differing but covariant coordinate systems. But if an ``event'' is the occurrence of an outcome, the inconsistency of the counterexample is an inconsistency concerning the set of events themselves, so that it is not a benign matter of differing descriptions of the same set of events, as in relativity theory. Rather, we get an inconsistency concerning the set of events themselves, which is supposed to be the only reality in RQM. Thus, RQM's putative basic reality cannot exist. On the other hand, if RQM denies that outcomes constitute events, and ``event'' is just another word for a unitary interaction, its basic postulate is still falsified by the counterexample. 

\section{Does Everett survive?}

The present author is well aware that there are many and diverse variants and approaches to the Everettian picture, also called the ``Many Worlds Interpretation'' (MWI). (For a recent review and discussion, see  Waegell and McQueen, 2000). Although presentations of MWI typically represent ``measurement'' as involving human observers modeled by multiple degrees of freedom comprising ``sense organs'' and ``brains'' or ``memories,'' MWI implicitly adopts the UOA since it holds that there is no special interaction that qualifies as a ``measurement.'' This was certainly the original approach of Everett (1957) in his Relative State interpretation. MWI has been subject to criticism in that the basis for defining the relative states, even if well-defined in a particular instance (i.e., not subject to UOA degeneracy), does not necessarily correspond to localized, quasi-classical properties. For example, a correlation could be established defining relative momentum states, so that an experimenter would have to ``see'' a completely nonlocalized, wavelike property. The response to this problem has been to appeal to decoherence (we return to this issue below). 

In any case, if an Everettian grants that an experiment such as that of Proietti {\it et al} (2019) can confer outcomes upon correlated degrees of freedom such as photons or spin-1/2 systems, then he must hold that branching of worlds occurs at the level of the those ostensible ``observer'' photons playing the role of $F$. If it is assumed that the branching propagates at speed $c$ to the outside observer $W$, then $W$ must experience the outcome-defined states rather than the entangled state. But then we would have a conflict with experiment, since Proietti {\it et al} obtain data showing the persistence of the entangled state (not to mention the innumerable EPR-type experiments already done with such entangled states). Therefore, we must conclude either that (1) the entanglement of a new degree of freedom with another degree of freedom is not sufficient to constitute ``measurement'' under MWI or (2) the branching is confined to the level of the putative ``observer'' photons. 

If they go with (1), then Everettians must reject the claims of Proietti {\it et al} (2019) and any similar proposal that asserts that outcomes occur for microscopic subsystems of correlated entangled states. This constitutes a form of ``moving the goalpost'' to some desired level of ``macroscopicity,'' which introduces an {\it ad hoc} element (see also the discussion at the end of Section 2). It is essentially the same move as appealing to decoherence, and thus inherits the same ambiguities and limitations (see below). On the other hand, if an Everettian goes with (2) and allows that Proietti {\it et al} are correct to attribute outcomes to entangled photons, then they are subject the counterexample as follows. 

 MWI posits two distinct worlds $U$ and $D$ corresponding to $F$'s possible ``up'' or ``down'' outcomes, and two corresponding versions of $F$ himself: $F^U$ and $F^D$, each of which finds a sharp value for the relevant spin observable. But if we assume that $W$ cannot have branched since, in conformance with experiments, he can confirm this sort of entangled state, when W measures $F$ and $A$, he finds information corresponding to a sharp value of a non-commuting observable (i.e., a confirmation of the $A$'s prepared state), and communicates that information to both versions of $F$ (through their shared degree of freedom $C$).\footnote{That is, if it makes ontological sense for two distinct versions of $F$ to share a single degree of freedom in this manner, which could be questioned. If this is not possible, then MWI breaks down right away, since $W$ is supposed to send a single photon to $C$ in each run of the experiment, but there are now two versions of $C$. } Thus, $F$ has information about sharp values for non-commuting observables for the same system, in violation of the uncertainty principle; or, if one objects to the uncertainty principle violation based on UOA degeneracy (see footnote 7), then $F$ has information about both (i) a highly sharp and (ii) completely uncertain value for $\sigma_z$ (since W's state assignment places $A$ in an improper mixed state), also a contradiction. This problem is not remedied by $F$'s simply concluding that his state assignment must be wrong and updating it in light of new information from $W$, because it is claimed that $F$ really did obtain an outcome for his putative measurement.\footnote{Baumann and Brukner (2019) mention Everettian (Many World Interpretations or MWI) as a way of evading/accommodating the inconsistency by simply allowing $F$ to update his state assignment based on the new information from $W$. Besides the FAPP nature of such a move, they don't appear to take into account the inconsistency concerning the information about $\sigma_z$.}
 
Can an Everettian avoid both the above problems (relative states not classical, uncertainty principle violation or inconsistent information about $F$'s observable) by rejecting UOA and asserting that outcomes occur only for a situation involving significant decoherence? First, he must specify under what circumstances outcomes {\it do} occur. Since the advent of standard unitary-only decoherence is not a digital but rather a virtually continuous process (in which off-diagonal terms of the reduced density matrix steadily diminish), this would require an arbitrary `cut' at which point it is stipulated that an outcome occurs. Besides the obvious {\it ad hoc} nature of this move (not to mention the circularity of the SQT decoherence program, Kastner (2014, 2020a)), the inconsistency identified and unresolved by Laudisa and Rovelli remains: however small, interference effects are supposedly still present according to SQT after the cat has `measured' its heartbeat (after $F$ has `measured' $A$). This means that ``branching'' cannot be absolute; there must be some ``interference'' of the branches. The external system ($W$) could in-principle detect this interference and would then send a photon to $F$ in accordance with the relevant probabilities for $W$'s measurement, allowing for the effects of a significant amount of decoherence to reduce the visibility of the Bell observable measurement. However, even if the visibility of the Bell observable measurement is very low, we would still have an inconsistency, since it is asserted that $F$ has found a sharp value of an incompatible observable. One might object that such an experiment would be impractical, but that does not remove the in-principle inconsistency.

\section {Consistent Histories Inadequate}

Finally, I address what in my view is the inadequate treatment of this scenario by the Consistent Histories (CH) approach, formulated by Robert Griffiths (1984). CH was developed to try to provide an account of the observed classicality of the macroscopic world despite the quantum measurement problem. CH identifies classically separable histories based on the multiple-time compatibility of the sequence of observables and/or states making up a history.  

 Griffiths describes this program as based on so-called ``Hilbert Space Quantum Mechanics,'' which includes von Neumann's form of the projection postulate---``Process 1''---involving the Born Rule (Griffiths 2021, p. 2). In this sense, CH does not adopt the UOA, since it reserves the right to include a stochastic interlude prior to a putative measurement outcome or possessed property, and indeed it purports to solve the measurement problem by helping itself to the projection postulate whenever such a determinate property is desired. However, as noted previously, Process 1 is an {\it hoc} postulate rather than a fundamental physical principle.\footnote{Indeed the Born Rule, under SQT, is an instrumentalist recipe arrived at as an afterthought by Born when he noted (in a footnote of Born, 1926) that one needed to square the wavefunction in order to obtain a probability. The {\it ad hoc} nature of the Born Rule under SQT becomes quite clear if one considers the actual history of the arrival of the physics community at this rule, as related by A. Pais (1982): ``...Born originally associated probability with $\Phi_{mn}$ rather than $|\Phi_{mn}|^2$. As I learned from recent private discussions, Dirac had this very same idea at that time. So did Wigner, who told me that some sort of probability interpretation was then on the minds of several people, and that he, too, had thought of identifying probability with $\Phi_{mn}$ or $|\Phi_{mn}|$. When Born's paper came out and $|\Phi_{mn}|^2$  turned out to be the relevant quantity, `I was at first taken aback but soon realized that Born was right,' Wigner said.'' (Pais 1982, p. 9). On the other hand, the Transactional Interpretation derives the Born Rule and its accompanying non-unitarity from physical principles. See, e.g., Kastner (2022, Chapters 3,5) and Kastner (2021a).} Von Neumann understood that his ``Process 1'' had no physical counterpart in standard quantum theory, and he could only think of it as somehow requiring the consciousness of an external observer, involving a ``split'' between the system(s) described by quantum theory and some ``external'' system. The whole point of the Wigner's Friend paradox is that it precludes appealing to such ``externality.'' Thus, CH amounts to a form of instrumentalism rather than an account of physical processes in the world. We return to this point below, but first let us briefly review the CH formulation. 
 
 The Hilbert space $\breve{H}$ for a history $Y$ comprising various events $n$ for a sequence of times $t_n$ is defined as a multiple-time copy of the Hilbert space for the system, i.e.:

$$\breve{H} = H_1 \odot H_2 \odot H_3 \odot  \dots  \odot H_N \eqno(4)$$

A specific history $Y$ under CH is a sequence of events described either by a state preparation, a putative possessed property, or a measurement outcome. Each of these is represented by a corresponding projector denoted by $[ P ]$. For example, for a typical Stern-Gerlach measurement, the history of the system prepared in the state $|\Psi_0\rangle$ could be represented by: 
 
$$Y = [ \Psi_0 ] \odot [ \Psi_1 ] \odot [Z \uparrow ] \eqno(5),$$

\noindent where $ |\Psi_1\rangle  = U(t_1 - t_0) |\Psi_0\rangle $ and the system has been found in the outcome ``up along Z''.

A family of histories, or ``framework,'' is generated from a ``projective decomposition of the identity'' (PDI) where the set of projectors at each event is orthogonal and complete. The framework corresponding to the history $Y$ with the given preparation would be expressed as

$$  F_Y = [ \Psi_0 ] \odot \{[ \Psi_1 ], [ 1-  \Psi_1 ]\} \odot\{ [Z \uparrow ], [Z \downarrow ] \} \eqno(6)$$

Based on this framework, we could identify two histories of interest corresponding to the two possible outcomes of the 
final measurement:

$$Y_{\uparrow} = [ \Psi_0 ] \odot [ \Psi_1 ] \odot [Z \uparrow ] \eqno(7a)$$
and
$$Y_{\downarrow} = [ \Psi_0 ] \odot [ \Psi_1 ] \odot [Z \downarrow ] \eqno(7b)$$

In order for the probabilities of these two histories to be additive, i.e.

$$P(Y_{\uparrow} \lor Y_{\downarrow}) = P(Y_{\uparrow}) + P(Y_{\downarrow}) \eqno(8)$$

\noindent their framework must satisfy a consistency condition that is a generalization of the Born Rule. This boils down to

$$Tr [{Y^{\uparrow}}^\dag Y_{\downarrow}] =0 \eqno(9),$$

\noindent which is readily satisfied by the framework (6).

Now let us consider the history corresponding to the counterexample. Suppose that $A$ is prepared in the state 
$|\psi\rangle_0 = \frac{1}{\sqrt 2} [|z\uparrow\rangle + |z\downarrow\rangle]$
at $t_0$, $F$'s measurement takes place at $t_1$, $W$'s measurement takes place at $t_2$, and W sends his photon to $F$ at $t_3$.  The first thing to note is that $F$ and W must disagree on the decomposition of the identity $I_1$ applying at $t_1$:  $F$ thinks it is

$$ {I_1}^F \rightarrow \{  [Z \uparrow \otimes B \uparrow], [Z \uparrow \otimes B \downarrow], [Z \downarrow \otimes B \downarrow] ,[Z \downarrow \otimes B \uparrow] \}  \eqno(10)$$

\noindent while $W$ thinks the decomposition is:

$$ {I_1}^W \rightarrow \{  [\Phi^+ ],[1-\Phi^+  ]  \} \eqno(11)$$

The decomposition of the multi-time identity $\breve{I}$ ceases to be well-defined at that point. Their decompositions of the identity at $t_2$, when $W$ measures the Bell observable, must also disagree. Thus, there is not only no consistent framework for the situation, but not even a well-defined inconsistent framework, since $F$ and W disagree on the decomposition itself!

The lack of any well-defined (let alone consistent) framework for the case in which both observers $F$ and $W$ obtain outcomes under SQT's presumed persistence of unitary evolution is a nice way to see the logical nature of the inconsistency (though I will argue below that the CH diagnosis of the problem is not a solution of it). If $F$ is assumed to find an outcome of either `up along z' or `down along z',  those have to count as projectors in the history, and therefore as a basis in the decomposition of the multi-time identity $\breve{I}$ (as in equation (10).  But based on SQT's persistence of unitarity, $W$ must apply an incompatible decomposition of the identity at the {\it same time}--one with respect to the Bell-state basis (of which $\Phi^+$ is one of the eigenstates)---as in equation (11). 

 Thus, in order to apply a consistent framework, CH has two main alternatives: (1) It must deny that $F$ finds an outcome at $t_1$ (pertaining to the decomposition (10) while $W$ does find an outcome at time $t_2$ based on the putative persistence of the state $|\Phi^+\rangle$ and his measurement of the Bell observable. Yet this amounts to a double standard: if all systems are undergoing only unitary evolution (the SQT paradigm), why should $F$ {\it not} get an outcome (in accordance with unitary evolution) while W {\it does} (in violation of the unitary evolution)? Moreover, if $F$ were a human scientist (as in the example of Baumann and Brukner (2019), CH would have to deny that human scientists find outcomes for their measurements, which of course is unacceptable. 
 
This leads to the second alternative: (2) allowing $F$ to find an outcome and denying that $W$ is working with the state
  $|\Phi^+\rangle$, but instead is working with what amounts to a proper mixed state based on $F$'s possible outcomes---in short, to apply the Projection Postulate (PP)  ``across the board'' at time $t_1$.  But if $F$ is only a molecule or other microscopic quantum system subject to ongoing entanglement, clearly we would not want $F$ to find an outcome. Thus, we again have a double standard concerning to what and for whom the PP applies. An attempt to escape this double standard would take the form of an appeal to decoherence, but this founders on the same arbitrariness problem mentioned earlier (no principled way to define the onset of outcomes; helping ourselves to an outcome when we want it, even though SQT itself does not provide that).
  
At this point, a few general criticisms of the CH program are in order (see also Kent (1997) and Okon and Sudarsky (2015)). 
The CH recipe is to help oneself to an unexplained ``stochastic'' interlude (basically the PP with no accompanying physics) prior to the desired measurement outcome or putative property assignment. Thus, rather than a solution to the measurement problem, it amounts to an evasion. As long as no quantitively specified alternative physical evolution is provided, unitary evolution still applies to the system, whether or not one chooses to ignore it. Physical reality is not affected by a physicist's representational choices, as Griffiths himself emphasizes: ``The choice of which framework to use is made by the physicist when analyzing experimental data in order to understand its physical significance, and has no influence on the actual physical process.'' (Griffiths 2021, p. 4)\footnote{Although arguably this does not lead to ``understanding'' of what might be going on, but rather to the sort of story that one wishes to have, whether or not quantum theory actually provides it. This is emphasized by the mutually inconsistent representations constructed for a Stern-Gerlach measurement in Griffiths (2015, Section 5), only one of which (equation 10) corresponds to the actual unitary evolution of the system as required by SQT, which Griffiths acknowledges. The others (eqs. (11), (12), and (15)) give mutually conflicting stories about properties that {\it would} be possessed by the system {\it if}, contrary to SQT, there had been real non-unitarity at different times, and are therefore completely fictitious as accounts of what the system(s) are actually doing under SQT.  The plethora of mutually inconsistent accounts of a single physical process is referred to as a violation of ``unicity'' under CH, and it is claimed that unicity should be abandoned. But again, mutually incompatible stories conceived by an experimenter for his convenience are no more than thoughts, and as such have nothing to do with what the physical system is doing, especially if the underlying theory quantitatively provides only unitary evolution.} So helping oneself to a physically unexplained stochastic transition yielding a desired measurement basis is inconsistent with the unitary evolution as demanded by SQT.\footnote{Griffiths says: ``Thus the first measurement problem is resolved in the CH approach by employing a sample space that includes the final measurement outcomes (pointer positions) in its description, together with the unitary time development generated by the standard Schr\"{o}dinger equation to assign probabilities using Born's formula.'' (Griffiths, 2015). Besides the obvious instrumentalism of this account, this is very much like saying: ``The problem of arriving in Spain after having boarded a plane apparently bound for Alaska is resolved by employing a final destination space that includes the final destination of Spain in its description.''  Again, the only quantifiable physical evolution available to the CH program is unitary, and the Born Rule is used as an {\it ad hoc} instrument.  A {\it description} of observed phenomena is neither an {\it explanation} of those phenomena nor a resolution of an inconsistency between theory and phenomena.}  

Moreover, CH allows the application to the same physical situation of many different frameworks calling for postulated stochastic transitions at conflicting times (see footnote 10). Those different descriptions deviate {\it physically} not only from the quantitatively prescribed purely unitary evolution of the standard theory but also from one another. CH thus helps itself to varying {\it ad hoc} representations to obtain varying (and mutually inconsistent) narratives portraying the existence of outcomes or desired properties where the assumed unitary dynamics does not provide any of these.  That is, as long as the theory in play is SQT (meaning a theory lacking quantifiable, real non-unitary physical behavior under clearly specified conditions), it does not lead to any of the desired frameworks representing the existence of outcomes, and they are all fictions. CH is thus used as a tool not only for ``saving the appearances'' but also for telling ourselves whatever mutually conflicting stories we might desire about the possession of properties at various times, regardless of what the underlying theory physically demands.
 
Finally, it should be noted that Losada et al (2019) provide a CH analysis of the Frauchiger-Renner scenario. They comment that the putative outcomes and resulting predictions at the Friend-level and the Wigner-level belong to mutually inconsistent histories. (Their construction applies the relevant outcome-related decompositions at different times). Their conclusion based on this construction is to say that the outcomes should not be combined in what they term a `classical conjunction,' and that doing so is a mistake at the level of theoretical inference. That is, they appear to assume that the F-R scenario, and by implication the Wigner's Friend scenario, is non-problematic for SQT in that the conflicting outcomes would only obtain privately in the individual contexts of the Friend(s) and Wigner(s). However, the counterexample discussed herein demonstrates that one does not need an illicit classical conjunction in order for the inconsistency to become manifest. For the systems involved are quantum systems undergoing only quantum evolution, and the inconsistency does not depend on any sort of classical record or inference. Thus, it is SQT itself that leads to the inconsistency. This is the same point made in section 2.

\section{Conclusion}

I have discussed an example showing that Standard Quantum Theory (SQT), together with the Unitary Outcome Assumption (that outcomes obtain based only on unitary correlations), can lead to an empirically consequential inconsistency. The inconsistency takes the form of two mutually exclusive values for the probability of a local change in the inner ``observer'' system $F$'s state. Its further implications consist either of (1) a violation of the uncertainty principle or (2) a contradiction concerning $F$'s result. Specifically: (1) according to the UOA, $F$ obtained a sharp value of  $\sigma_z$ for $A$, which should make $A$'s value of $\sigma_x$ maximally uncertain, but $W$'s measurement outcome allows the inference of a precise value for $\sigma_x$; or (2), as in footnote 4 taking into account degeneracy of the state $|\Phi^+\rangle$ with respect to the individual spin basis, $F$ has information about both a maximally sharp {\it and} maximally unsharp value of  $\sigma_z$ for $A$ (or, allowing for significant decoherence, a maximally sharp {\it and} unsharp $\sigma_z$ value). Neither of these contradictions can be eliminated by $F$'s updating his state assignment, since he supposedly obtained a sharp measurement outcome for $\sigma_z$, which is contradicted by $W$'s result communicated to him. The inconsistency cannot be avoided by hidden variables either, unless under such a theory an experimenter declines to assign an outcome-eigenstate to his measured/prepared system (i.e., declines to apply quantum theory).

However, keeping SQT while rejecting UOA is hardly satisfactory, since then the condition for obtaining an outcome cannot be defined in any consistent way. Appealing to decoherence is not sufficient, since the onset of decoherence is essentially continuous and the occurrence of a definite outcome is digital, making the degree of decoherence deemed sufficient to ``trigger'' an outcome an arbitrary matter. In particular, under MWI the existence of an outcome demands splitting: we either have splitting or we don't. In any case, as noted above, an inconsistency remains as long as any non-vanishing coherence exists. Thus, SQT appears fatally inconsistent, at least from a realist (and non-FAPP) stance. The finding is not limited to single-outcome interpretations but also infects hidden variables (including ``superdeterministic'' theories), Everettian (MWI) approaches, and Relational Quantum Theory. Instrumentalist approaches to SQT are still working with a flawed instrument, and their state assignments must generally be viewed as placeholders subject to correction given new information; hence they are afflicted by an inevitably FAPP, quantitatively ill-defined nature. It has also been argued that Consistent Histories fails to address the problem, since it amounts to a form of arbitrary instrumentalism that simply suspends unitary evolution at any desired point or points while giving no quantitative account of any real non-unitary process. Thus the different history ``options'' yielding measurement outcomes or putative properties under CH are physically unjustified as long as the underlying theory provides only unitary evolution. These representations do not represent something in the world, but are fictions. 
 
Baumann and Brukner (2019) note that spontaneous collapse approaches are not subject to the inconsistency. However, they assume that all such models involve {\it ad hoc} changes to quantum theory resulting in ``new physics.'' In this regard, they take GRW-type approaches as the only sort of collapse theory, which is not an accurate portrayal of the state of research. Their account therefore presents a false choice: it implies that one must either (i) stick with the standard theory and be subject to its inconsistencies or (ii) settle for an explicitly {\it ad hoc} alteration of quantum theory.  But in fact, this typical portrayal of the `state of play' fails to accurately represent our available theoretical options. While the Penrose model does result in empirical deviations from standard quantum theory, it differs from GRW-type approaches in that it does not incorporate {\it ad hoc} changes to the Schr\"{o}dinger equation but appeals to another established theory (gravitation) in order to effect reduction. Meanwhile, the Transactional Interpretation (extended into the relativistic domain by the present author and called RTI), yields predictions that are fully empirically equivalent to the standard theory at the level of the Born Rule (this is a theorem; for details, see Kastner (2021a)). Thus, the ``new physics'' of TI/RTI is not really new, but is simply that of the direct-action theory of fields, which allows derivation of the Born Rule based on physical principles. The only empirical deviation of RTI from the standard theory (SQT) is that RTI predicts determinate outcomes with real non-unitary reduction to the corresponding eigenstates under quantitative, physically specified conditions (e.g., Kastner, 2018), while SQT does not. In other words, RTI predicts that if $F$ gets an outcome, it is because there has been a non-unitary interaction placing his system in a proper mixed state, and $W$'s measurement will confirm this (i.e., entanglement is broken at the level of $F$). For further details, the interested reader may consult Kastner (2022), (2018) and Kastner and Cramer (2018), Kastner (2021a), (2022). 

It is sometimes viewed as a drawback that the transactional formulation does not yield testable ``new predictions''; i.e., deviations from the Born Rule. However it should be recalled that the current theory faces empirical-level anomalies such as the inconsistencies discussed herein. RTI passes these empirical tests by predicting results in conformance with what is observed (e.g., determinate measurement results and no inconsistencies), thus resolving both empirical and theoretical anomalies for the standard theory. This situation is analogous to the anomaly of the advance of Mercury's perihelion, which was resolved by relativity theory. This was not a ``new prediction'' per se, but rather a prediction of existing empirical data that could not be accommodated by the current theory. Thus, the ability of a new theory or formulation to predict what is  anomalous for the current theory (e.g., single-outcome measurement results that form a consistent set) is a robust form of confirmation for the competitor over its predecessor.

Finally, the present author is in full agreement with Sir Roger concerning the need to embrace non-locality as a fundamental feature of the quantum level, a view he expressed in his book {\it The Large, the Small, and the Human Mind}: 

``My own view is that to understand quantum non-locality we
shall require a radically new theory. This new theory will not
just be a slight modification to quantum mechanics but something as different from standard quantum mechanics as
General Relativity is different from Newtonian Gravity. It
would have to be something which has a completely different
conceptual framework. In this picture, quantum non-locality
would be built into the theory.'' (Penrose, 1997).

Both our approaches to a quantum theory could perhaps be considered this sort of ``radical'' retooling of quantum theory. The different conceptual framework in the transactional approach concerns the mutual and reciprocal behavior of fields, which generates non-unitarity, while in Sir Roger's approach the new conceptual framework is to attribute to Nature the ability to generate reduction in order to obtain a well-defined metric. In both cases it is allowed that the world is nonlocal at a fundamental level, and that this should be embraced rather than evaded.

A further observation is perhaps in order from a broader perspective, that of considerations of general theory evaluation in science and in particular the assessment of anomalies facing established theories. L. Gradowski (2022) has made a cogent case that adherents of established theories tend to minimize or dismiss anomalies for those theories, and to oppose competing approaches.\footnote{Gradowski observes that theories often initially thought of as ``fringe'' often are later vindicated and become established theories.} Formulations of quantum theory involving physical reduction are generally viewed in the latter (dim) light, while the measurement problem facing the standard theory is generally not acknowledged as the serious anomaly that it is. In a nutshell, empirically we find single outcomes; the standard theory fails to predict those single outcomes. As discussed herein, even when augmented by {\it ad hoc} strategies to force single outcomes, such as the projection postulate or hidden variables, the standard theory is still subject to inconsistencies. 

One manifestation of anomaly avoidance on the part of adherents of standard quantum theory has been to simply stipulate (on the basis of the Unitary Outcome Assumption) that single outcomes occur and can be mutually inconsistent. This paper has  observed, however, that such inconsistencies cannot be counted on to remain private and incommensurable. Instead, they rise to the level of empirical inconsistency, such that it is the theory itself that is self-contradictory: a single observer could in-principle be subject to mutually exclusive predictions {\it by the theory itself} (not by any illicit inference on an observer's part). This is a failure of the theory, not with anyone's interpretation or use of it. Yet it has still not been fully recognized as the ``nail in the coffin'' of the standard theory, arguably due to theoretical entrenchment as delineated in Gradowski's insightful and historically-informed study of theory evaluation. Gradowski notes:

``Theoretical entrenchment is not simply a matter of conservatism, which may sometimes be healthy for a science. Rather, it invokes biases that come with investment in a theory, such as closed-mindedness and epistemic insensitivity to relevant and promising alternatives. Theory-laden interpretations of evidence and neglect of evidence can both be seen as negative results of theoretical entrenchment. Furthermore, we would like to believe that experts are fair-minded, but evidence indicates that this may not be the case when core views are threatened.'' (2022, p. 6).

This situation invites us to recall (as noted in Kastner 2020b) that the original intent of Schr\"{o}dinger's Cat Paradox was to highlight for critical examination an anomaly afflicting the standard theory: the measurement problem; i.e., the inability of the theory to account for measurement outcomes. Wigner accentuated the anomaly with his ``Friend'' version of the paradox. Yet, rather than recognize and address the {\it reductio} nature of these scenarios as manifestations of a critical problem with the standard theory, the physics and philosophy of physics communities have (in general) systematically elevated these anomalies to putatively interesting and instructive ``lessons'' about the theory and/or about the world.\footnote{A key example of theory-laden dressing up of anomalies as `lessons' or `new information' is the so-called `violation of unicity' invoked by Griffiths to describe the Consistent Histories expedient of telling ourselves whatever story helps us save our immediate empirical appearances, even when the end result is several mutually inconsistent stories. (see note 14). In a more recent preprint (Griffiths 2023), he uses the term `violation of {\it classical} unicity,' implying that expecting a coherent and logically self-consistent account of the physical world that accommodates our empirical observations is a form of classical prejudice or naivete. Going this route gives license to any future theory to be self-inconsistent and incoherent based on the theory-laden claim that the quantum level `violates unicity' (when the current theory cannot even coherently explain what counts as a `quantum system' as opposed to a `classical system'). Moreover, it flouts a basic standard of logical inference: the fundamental requirement of self-consistency, by essentially demoting it to a classical prejudice that should be laid down. This highlights the danger in theoretical entrenchment: rather than question the theory yielding the inconsistencies, basic scientific and logical principles are jettisoned in order to recolor its anomalies as acceptable or even as `lessons'. The situation is arguably pathological.} 

  This can be seen as an example of the theory-laden interpretation of evidence (or problematic predictions of a theory): rather than recognize the predictions as problematic, the entrenched theory that yields them is not questioned and instead an interpretation is advanced in which problematic and even contradictory predictions are to be accepted as inevitable. This is the situation obtaining in treatments of the Wigner's Friend-type inconsistencies that eschew rejecting the standard theory and instead insist that its inconsistent predictions obtain ``subjectively'' or in a locally observer-dependent way. Yet of course, that approach fails in the light of the counterexample discussed herein.

In conclusion, a theory that is empirically insufficient (fails to predict observed single outcomes) and even self-inconsistent (yields empirical-level inconsistencies) cannot sensibly be viewed as either successful or instructive. Therefore, I believe that it is long past time to seriously consider the competitor (non-standard) theories that offer fruitful solutions to the fatal problems afflicting the standard theory. Among those are the Di\'{o}si-Penrose gravitational collapse theory and the Relativistic Transactional Interpretation. These have the advantage (over the GRW approach) of either no change to the Schr\"{o}dinger Equation (for the case of RTI) or a change that is motivated by gravitation, another existing theory of Nature.

\newpage

\section{References}

Andreoletti, G., Vervoort, L. Superdeterminism: a reappraisal. {\it Synthese 200}, 361 (2022). https://doi.org/10.1007/s11229-022-03832-6.

Baumann, V. and Wolf, S. (2018). ``On Formalisms and Interpretations,'' {\it Quantum 2}, 99.

Baumann, V. and Brukner, C. (2019). ``Wigner?s friend as a rational agent,'' preprint, https://arxiv.org/pdf/1901.11274.pdf

Bong, K.-W. et al (2020). ``A strong no-go theorem on the Wigner's friend paradox,'' {\it Nature Physics, Vol.16}, 1199-1205.

Born, Max. (1926). ``Zur Quantenmechanik der Stossvorg \''{a}nge,'' {\it Zeitschrift fur Physik 37}, 863-867.

Brukner, C.(2015).  ``On the quantum measurement problem,'' in Proceedings of the
Conference ``Quantum UnSpeakables II: 50 Years of Bell?s Theorem'' ; 
arXiv:1507.05255.

Clauser, J. M. Horne, A. Shimony, and R. Holt (1969). {\it Phys.Rev. Lett.23}, 880.

Cramer, J. G. (1986). Rev. Mod. Phys. 

D. Deutsch, (1985). Quantum theory as a universal physical theory, Int. J. Th.
Phys. 24, I.

Di\'{o}si, L. (1987). "A universal master equation for the gravitational violation of quantum mechanics". Physics Letters A. 120 (8): 377?381. 

Di\'{o}si, L. (1989). "Models for universal reduction of macroscopic quantum fluctuations". Physical Review A. 40 (3): 1165?1174.

Di\'{o}si, Lajos (2013). "Gravity-related wave function collapse: mass density resolution". Journal of Physics: Conference Series. 442 (1): 012001.

Elouard {\it et al} (2021). ``Quantum erasing the memory of Wigner's friend,'' Quantum 5, 498.

Everett, H. (1957). ``Relative State Formulation of Quantum Mechanics'', {\it Reviews of Modern Physics 29}, 454?462.

Feyerabend, P. K.  (1962). ``On the Quantum Theory of Measurement.'' In Korner S. (ed.) (1962). {\it Observation and Interpretation in the Philosophy of Physics}. New York: Dover, pp. 121-130.

Frauchiger, D. and Renner, R. (2018). ``Quantum theory cannot consistently describe the use of itself,'' {\it Nature Communications 9}, Article number: 3711.

Gradowski, Laura (2022). ``Facing the Fringe''. CUNY Academic Works. https://academicworks.cuny.edu/gc\_etds/5091. 

Ghirardi,	G.C., 	Rimini,	A.,	and	Weber,	T.	(1986).	{\it Phys.	Rev.	D	34},	470.

Griffiths, R. (2015). ``Consistent Quantum Measurements,'' preprint: https://arxiv.org/abs/1501.04813.

Griffiths, R.  (2021). ``Bell nonlocality versus quantum physics:  A reply toLambare,'' preprint: https://arxiv.org/abs/2106.09824.

Griffiths, R. (2023).  "Consistent Quantum Causes", arXiv:2303.13617.

Hughes, R. I. G. (1989). {\it The Structure and Interpretation of Quantum Mechanics}. Cambridge: Harvard University Press. 

Kastner, R. E. (2022). {\it The Transactional Interpretation of Quantum Mechanics: A Relativistic Treatment.} Cambridge: Cambridge University Press.

Kastner, R. E. (2014). `` 'Einselection' of pointer observables: the new H-theorem?'' {\it Stud. Hist. Philos. Mod. Phys. 48}, 56-8.

Kastner, R. E.  (2018). ``On the Status of the Measurement Problem: Recalling the Relativistic Transactional Interpretation,'' {\it Int'l Jour. Quan. Foundations 4}, 1:128-141.

Kastner, R. E. (2020a). ``Decoherence in the Transactional Interpretation,'' {\it Int'l Jour. Quan. Foundations 6}, 2:24-39. 

Kastner, R. E. (2020b). ``Unitary-Only Quantum Theory Cannot Consistently Describe the Use of Itself: On the Frauchiger-Renner Paradox,'' {\it Foundations of Physics 50}, 441-456.

Kastner, R. E. (2021a). ``The Relativistic Transactional Interpretation and The Quantum Direct-Action Theory,'' preprint. https://arxiv.org/abs/2101.00712. (This material is excerpted from Kastner (2022, Chapter 5).

Kastner, R. E. (2021b). ``Unitary Interactions Do Not Yield Outcomes: Attempting to Model `Wigner?s Friend','' {\it Found Phys 51}, 89 (2021).
doi:10.1007.s10701-021-00492-3

Kent, A. (1997). ``Consistent sets yield contrary inferences in quantum theory,'' {\it Phys. Rev.Let.}, vol. 87:15.

Laudisa, Federico and Carlo Rovelli (2021). "Relational Quantum Mechanics", The Stanford Encyclopedia of Philosophy (Spring 2021 Edition), Edward N. Zalta (ed.), URL = <https://plato.stanford.edu/archives/spr2021/entries/qm-relational/>. 

Losada, M., Laura, R., and Lombardi, O. (2019). ``Frauchiger-Renner argument and quantum histories,'' {\it Phys. Rev. A 100}, 052114.

Matzkin, A. and Solokovski (2020). ``Wigner?s friend, Feynman?s paths and material records,'' {\it EPL 131}, 40001.

Okon, E.  and Sudarsky, D (2015). ``The Consistent Histories Formalism and the Measurement Problem,'' {\it Stud. Hist. Phil. Mod. Phys. 52}, 217-222

Pais, A. (1982). ``Max Born's Statistical Interpretation of Quantum Mechanics.'' {\it Science 218}(4578), 1193-1198. http://www.jstor.org/stable/1688979

Penrose, R. 1997. {\it The Large, the Small and the Human Mind.} Cambridge University Press,
Cambridge.

Penrose, Roger (1996). "On Gravity's role in Quantum State Reduction". General Relativity and Gravitation. 28 (5): 581?600. 

Penrose, Roger (2014). "On the Gravitization of Quantum Mechanics 1: Quantum State Reduction". Foundations of Physics. 44 (5): 557?575.

Proietti, M. et al (2019). ``Experimental test of local observer independence,'' {\it Science Advances}, 
Vol. 5, no. 9. DOI: 10.1126/sciadv.aaw9832 

Rovelli, Carlo, 1996, ``Relational Quantum Mechanics,'' International Journal of Theoretical Physics, 35(8): 1637?1678. doi:10.1007/BF02302261.

Waegell, W. and McQueen, K. J. (2000). ``Reformulating Bell?s Theorem: The Search for a Truly Local Quantum Theory,'' {\it Studies in the History and Philosophy of Modern Physics 70}:39-50.

Wesley,	D.	and	Wheeler,	J.	A.	(2003). ``Towards	an	action-at-a-distance	concept	of	
spacetime,''	In	A.	Ashtekar	et	al,	eds.	(2003).	Revisiting the Foundations of Relativistic
Physics:	Festschrift in Honor of John Stachel, Boston	Studies	in	the	Philosophy	and	
History	of	Science	(Book	234),	 pp.	421-436.	Kluwer Academic	Publishers.	

Zukowski, M. and Markiewicz, M. (2021). ``Physics and Metaphysics of Wigner's Friends: Even Performed Premeasurements Have No Results,'' {\it Phys. Rev. Lett. 126} (13).

\end{document}